\documentclass
[twocolumn,showpacs,preprintnumbers,amsmath,amssymb,superscriptaddress,floatfix,nofootinbib]{revtex4}
\usepackage{graphicx}
\usepackage{amsmath}
\usepackage{amsfonts}
\usepackage{amssymb}%
\begin{document}

\newcommand{\Ima}{\textrm{Im}}
\newcommand{\Rea}{\textrm{Re}}
\newcommand{\mev}{\textrm{ MeV}}
\newcommand{\be}{\begin{equation}}
\newcommand{\ee}{\end{equation}}
\newcommand{\ba}{\begin{eqnarray}}
\newcommand{\ea}{\end{eqnarray}}
\newcommand{\gev}{\textrm{ GeV}}
\newcommand{\nn}{{\nonumber}}
\newcommand{\dtres}{d^{\hspace{0.1mm} 3}\hspace{-0.5mm}}
\newcommand{\rts}{ \sqrt s}
\newcommand{\non}{\nonumber \\[2mm]}

\title{The $DN$, $\pi \Sigma_c$ interaction in finite volume and the  $\Lambda_c(2595)$ resonance}

\author{Ju-Jun Xie}
\email{xiejujun@ihep.ac.cn} \affiliation{Department of Physics,
Zhengzhou University, Zhengzhou, Henan 450001, China} \affiliation{
Departamento de F\'{\i}sica Te\'orica and IFIC, Centro Mixto
Universidad de Valencia-CSIC, Institutos de Investigaci\'on de
Paterna, Aptdo. 22085, 46071 Valencia, Spain }

\author{E. Oset}
\email{oset@ific.uv.es} \affiliation{ Departamento de F\'{\i}sica
Te\'orica and IFIC, Centro Mixto Universidad de Valencia-CSIC,
Institutos de Investigaci\'on de Paterna, Aptdo. 22085, 46071
Valencia, Spain }

\begin{abstract}

In this work the interaction of the coupled channels $DN$ and $\pi
\Sigma_c$ in an SU(4) extrapolation of the chiral unitary theory,
where the $\Lambda_c(2595)$ resonance appears as dynamically
generated from that interaction, is extended to produce results in
finite volume. Energy levels in the finite box are evaluated and,
assuming that they would correspond to lattice results, the inverse
problem of determining the phase shifts in the infinite volume from
the lattice results is solved.  We observe that it is possible to
obtain accurate $\pi \Sigma_c$ phase shifts and the position of the
$\Lambda_c(2595)$ resonance, but it requires the explicit
consideration of the two coupled channels. We also observe that some
of the energy levels in the box are attached to the closed $DN$
channel, such that their use to induce the $\pi \Sigma_c$ phase
shifts via L\"uscher's formula leads to incorrect results.

\end{abstract}

 \maketitle


\section{Introduction}

One of the topics where efforts are recently devoted within Lattice
QCD is the  determination of hadron spectra, both in the meson and
baryon
sector~\cite{Nakahara:1999vy1,Nakahara:1999vy2,Mathur:2006bs,Basak:2007kj,Bulava:2010yg,
Morningstar:2010ae,Foley:2010te,Alford:2000mm,
Kunihiro:2003yj,Suganuma:2005ds1,Suganuma:2005ds2,Hart:2006ps1,Hart:2006ps2,
Wada:2007cp,Prelovsek:2010gm1,Prelovsek:2010gm2,Lin:2008pr,Gattringer:2008vj,Engel:2010my,Mahbub:2010me,Edwards:2011jj}.
After earlier claims of a successful determination of the hadron
spectra using rough approximations and large pion masses, work
continues along this line with more accurate approaches and problems
are arising that were not envisaged at first glance. The "avoided
level crossing" is usually taken as a signal of a resonance, but
this criteria has been shown insufficient for resonances with a
large width~\cite{Bernard:2007cm,Bernard:2008ax,misha}. The use of
L\"uscher's approach~\cite{luscher,Luscher:1990ux} is gradually
catching up. It is suited for the case when one has resonances with
one decay channel in order to produce phase shifts for this decay
channel from the discrete energy levels in the box. Yet, most of the
hadronic resonances have two or more decay channels and the need to
go beyond L\"uscher's approach becomes obvious. This method has been
recently improved in Ref.~\cite{misha} by keeping the full
relativistic two body propagator (L\"uscher's approach has an exact
imaginary part but makes approximations on the real part) and
extending the method to two, or more coupled channels, which had
also been addressed
before~\cite{Liu:2005kr,Lage:2009zv,Bernard:2010fp}, and continues
to catch the attention of the practitioners~\cite{sharpe,davoudi}.
The new method is also conceptually and technically simpler and
serves as a guideline for future lattice calculations. Continuation
of this new practical method have been done in
Ref.~\cite{mishajuelich} for the application of the J\"ulich
approach to meson baryon interaction and in Ref.~\cite{alberto} for
the interaction of the $DK$ and $\eta D_s$ system where the
$D_{s^*0}(2317)$ resonance is dynamically generated from the
interaction of these particles
\cite{Kolomeitsev:2003ac,Hofmann:2003je,Guo:2006fu,daniel}. The case
of the $\kappa$ resonance in the $K \pi$ channel is also addressed
along the lines of Ref.~\cite{misha} in Ref.~\cite{mishakappa}.

The investigation of different problems following
Refs.~\cite{misha,mishajuelich,alberto,mishakappa} is showing that
every case studied reveals particular features and there is no
common behaviour in the levels of the finite box nor on the way that
the phase shifts or bound states are obtained from the spectra of
the finite box. The study of these problems along those lines is
most useful, since it sheds light on how to deal with results of QCD
lattice calculations, which precision is needed in the lattice
results to accomplish a desired accuracy in the phase shifts or
resonance pole positions, and which strategy is most useful to
gather lattice results from where the results in infinite volume can
be obtained with maximum accuracy.

The case we report here is one more in this line, showing as a novel
feature how in this case some of the low lying levels in the finite
box, for energies where the $\pi \Sigma_c$ channel is open and the
$DN$ one is closed, are tied  basically to the $DN$ channel, such
that the blind use of L\"uscher's approach would lead to unrealistic
phase shifts for the $\pi \Sigma_c$ channel and the resonance
position. We show the problems that arise in this analysis and
provide the two channel approach to solve them.  These results
should be most useful when QCD lattice results are produced to
describe the $\Lambda_c(2595)$ resonance.

This paper is organized as follows. In Sec. II we show the formalism
of the $DN$ and $\pi \Sigma_c$ interaction in infinite and finite
volume. In Sec. III, the inverse problem of getting the phase shifts
from two channels analysis is shown, while in Sec. IV the results
obtained by using one channel analysis are shown. Finally, a short
summary is given in Sec. V.

\section{Formalism} \label{sec:formalism}

\subsection{The $DN$, $\pi \Sigma_c$ interaction in infinite volume}

The $DN$ system, in collaboration with coupled channels, leads to
the formation of a meson-baryon composite state, the
$\Lambda_c(2595)$~\cite{hofmannnpa763,mizutaniprc74,tolosprc77,juanprd79}.
In this section, we briefly revisit the form of the $\pi \Sigma_c$
and $DN$ interactions from the chiral unitary approach. This will
allow us to review the general procedure of calculating meson-baryon
scattering amplitudes. In the chiral unitary approach the scattering
matrix in coupled channels is given by

\be T=[1-VG]^{-1}V \label{bse} \ee where $V$ is the matrix for the
transition potentials between the channels and $G$, a diagonal
matrix, is the loop function for intermediate $\pi \Sigma_c$ and
$DN$ states, which is defined as

\begin{equation}
G=i\int \frac{d^4
q}{(2\pi)^4}\frac{1}{q^2-m^2+i\epsilon}\frac{2M}{(P-q)^2-M^2+i\epsilon}\,
\label{loop}
\end{equation}
where $m$ and $M$ are the masses of the $\pi$ or $D$ meson and the
baryon $\Sigma_c$ or $N$, respectively. In the above equation, $P$
is the total incident momentum of the external meson-baryon system.

In the present problem we have two main channels, $\pi \Sigma_c$ and
$D N$. There are also other channels considered in
Refs.~\cite{mizutaniprc74,tolosprc77,juanprd79}, such as $\eta
\Lambda_c$, $K \Xi_c$, $K\Xi_c^{\prime}$, $D_s \Lambda$, and
$\eta^{\prime}\Lambda_c$, which play a negligible role to generate
dynamically the $\Lambda_c(2595)$ state, and are not considered
here.

We study only the $s$-wave interaction, hence, the transition
interaction (potential) for channel $i$ to $j$ reads, \ba V_{ij} &=&
-\frac{C_{ij}}{4f^2}(2\sqrt{s}-M_i-M_j)  \nonumber
\\ &&
\times \left(\frac{M_i+E_i}{2M_i}\right)^{1/2}
\left(\frac{M_j+E_j}{2M_j}\right)^{1/2}, \label{chiralpotential} \ea
where $f=93$ MeV is the pion decay constant, $E_i$ and $E_j$ are the
energy of incoming/outgoing baryon $\Sigma_c$ or $N$. The transition
coefficients $C_{ij}$ are symmetric with respect to the indices, and
also isospin-dependent. By naming the channels, 1 for $\pi \Sigma_c$
and 2 for $D N$, the coefficients $C_{ij}$ for the case of isospin
$I=0$ are~\cite{mizutaniprc74} \be C_{11} = 4, \hspace{0.5cm}
C_{12}=C_{21}=\frac{\sqrt{6}}{8},\hspace{0.5cm} C_{22}=3. \ee

The loop function $G$ can be regularized both with a cutoff
prescription or with dimensional regularization in terms of a
subtraction constant. Here we make use of the dimensional
regularization scheme. The expression for $G$ is then
\begin{widetext}
\begin{eqnarray}
\label{eq:g-function}
 \mbox G^D_i(s, m_i, M_i) &=&  \frac{2M_i}{(4 \pi)^2}
  \left\{
        a_i(\mu) + \log \frac{m_i^2}{\mu^2} +
        \frac{M_i^2 - m_i^2 + s}{2s} \log \frac{M_i^2}{m_i^2}
  \right.
  \\
     &+& \frac{Q_i(\rts)}{\rts}
    \left[
         \log \left(  s-(M_i^2-m_i^2) + 2 \rts Q_i(\rts) \right)
      +  \log \left(  s+(M_i^2-m_i^2) + 2 \rts Q_i(\rts) \right)
    \right.
  \nonumber \\
  &- &
  \Biggl.
    \left.
    \log \left( -s+(M_i^2-m_i^2) + 2 \rts Q_i(\rts) \right)
      - \log \left( -s-(M_i^2-m_i^2) + 2 \rts Q_i(\rts) \right)
    \right]
  \Biggr\},
  \nonumber
\end{eqnarray}
\end{widetext}
where $s=E^2$, with $E$ the energy of the system in the center of
mass frame, $Q_i$ the on shell momentum of the particles in the
channel, $\mu$ a regularization scale and $a_i(\mu)$ a subtraction
constant. The form of Eq.~(\ref{chiralpotential}) is adapted from
the light hadron sector to the charm sector using SU(4) symmetry but
reducing the strength of the diagrams with a heavy vector exchange
by the weight of its propagator. One should, in principle, not
expect a good SU(4) symmetry, and actually it is broken in this case
through the use of physical masses in the propagator. But in basic
vertices the symmetry works quite well (see an extensive review in
Ref.~\cite{danithesis} and references therein, and in section IV of
the paper~\cite{wu}). We thus follow this approach, as done in
Refs.~\cite{hofmannnpa763,mizutaniprc74,tolosprc77,juanprd79}. One
must bear in mind that uncertainties in the actual value of $V_{ij}$
in Eq.~(\ref{chiralpotential}) are taken into account, in the spirit
of the renormalization group, by means of the freedom in the
subtraction constants of $G$ in Eq.~(\ref{eq:g-function}), which are
adjusted such as to get the energy of the $\Lambda_c(2595)$ in the
right place. Note that the only parameter-dependent part of $G$ is
$a(\mu)+\text{ln}\frac{m^2_i}{\mu^2}$. Any change in $\mu$ is
reabsorbed by a change in $a(\mu)$ through
$a(\mu^{\prime})-a(\mu)=\text{ln}\frac{\mu^{\prime 2}}{\mu^2}$.

\begin{figure}[ptbh]
\begin{center}
\includegraphics[scale=0.4]{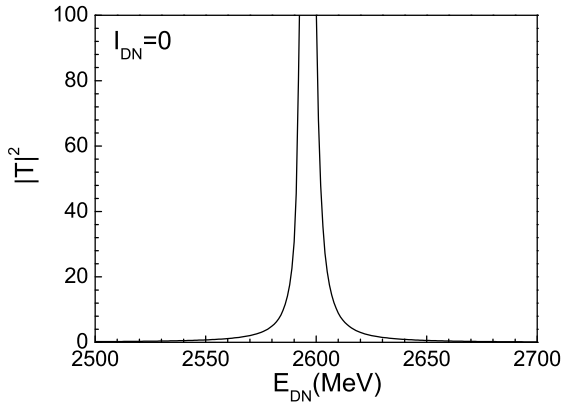}
\caption{Modulus squared of the $DN \to DN$ scattering amplitude for
$I_{DN}=0$.}
\label{Fig:tsDN}%
\end{center}
\end{figure}

In the infinite volume case, the use of Eq.~(\ref{bse}) with the two
channels that we consider leads to a dynamically generated state at
the energy of $2596$ MeV, which we associate to the
$\Lambda_c(2595)$ resonance, using the dimensionally regularized
$G^D$ function with $\mu=1000$ MeV and the subtraction constant
$a=-2.02$, which are values of natural size. In Fig.~\ref{Fig:tsDN}
the modulus squared of the $DN \to DN$ scattering amplitude as a
function of the invariant mass of the $D N$ system for $I_{DN}=0$ is
shown. There is a clear rather narrow peak around $2596$ MeV which
indicates the state of $\Lambda_c(2595)$ as a bound state of $DN$.

The poles of the amplitudes are found in the second Riemann sheet,
where in the loop function $G^D$ in Eq.~(\ref{eq:g-function})
$Q_i(\sqrt{s})$ is changed to $-Q_i(\sqrt{s})$ for the channel where
${\rm Re}\sqrt{s}$ is above threshold. The couplings of
$\Lambda_c(2595)$ to the $\pi \Sigma_c$ and $D N$ channels can be
obtained from the residues at the pole, by matching the amplitudes
to the expression \be T_{ij}=\frac{g_ig_j}{\sqrt{s}-\sqrt{s_R}} \ee
for $\sqrt{s}$ close to the pole $\sqrt{s_R}$. We find $\sqrt{s_R}=
(2596.3-i1.6)$ MeV. The couplings, $g_i$ and $g_j$ are complex in
general. In this way, we get $g^2_{\pi\Sigma_c}=0.65-i0.10$ and
$g^2_{DN}=43.3+i3.6$.

In Ref.~\cite{arriola}, an analytical study of bound states in
problems with coupled channels has been performed. A modern
formulation of the compositeness condition of
Weinberg~\cite{weinberg} for coupled channels has been derived from
a sum rule that comes from the normalization to unity of the wave
function of the bound state~\footnote{Technically, the state
obtained is a resonance because the $\pi \Sigma_c$ channel is open,
although with very small pase space. However, the sum rule also
holds at the resonant pole as shown in
Ref.~\cite{Sekihara:2010uz}.}, \be \sum_i g_i^2
\left.\frac{dG_{ii}}{dE}\right|_{E=E_\alpha}=-1, \label{eq76} \ee
where $g_i^2$ are the residues of the $T_{ii}$ scattering matrices
(coupling squared) at the pole  of the bound state ($\alpha$) and
$G_{ii}$ are the propagators of the two particles of the
corresponding channels (the loop function $G^D$ that we use here).

One can see from the derivation in Ref.~\cite{arriola} that each
term in Eq.~(\ref{eq76}) accounts (with reversed sign) for the
probability of the bound state to be made by the pair of particles
of the channel considered. By taking the coupling constant
$g^2_{DN}$ that we obtained above and the loop function $G^D_{DN}$,
we find that $g^2_{DN}\frac{dG^D_{DN}}{dE}=-0.91-i0.08$ at
$E=\sqrt{s_R}$, and, thus, about $91\%$ of the sum rule comes from
the $DN$ state, indicating that we have largely a bound $DN$ channel
in our approach.

\begin{figure}[ptbh]
\begin{center}
\includegraphics[scale=0.4]{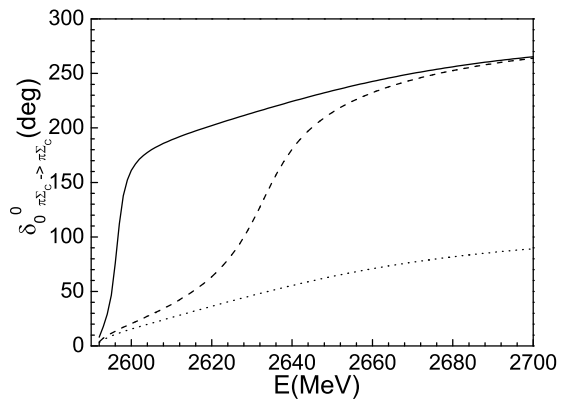}
\end{center}
\caption{Phase shift for $\pi \Sigma_c$ scattering derived from the
coupled channels unitary approach with different values of $\eta$.
Solid line: $\eta=1$. Dashed line: $\eta=0.8$. Dotted line: only
$\pi \Sigma_c$ channel considered.}
\label{Fig:pschiral}%
\end{figure}

Another way to check how important is the $DN$ channel in generating
dynamically of the $\Lambda_c(2595)$ state is to change the
parameters of the potential slightly, making the $\Lambda_c(2595)$
disappear. This can be achieved, for example, by merely reducing the
strength of the potential $V_{22}$ that describes the scattering in
the $DN \to DN$ channel. Namely, we replace $V_{22} \to \eta V_{22}$
and vary $\eta$ between 1 and 0, and as can be seen in
Fig.~\ref{Fig:pschiral}, the phase shifts for the $\pi \Sigma_c \to
\pi \Sigma_c$ scattering amplitude obtained from the Bethe-Salpeter
equation, Eq.~(\ref{bse}), drastically change with the strength of
$V_{22}$. The normalization that we use is such that in one channel
\cite{ollernpa620}~\footnote{We mention that in the present
calculation, we replace $T(E)$ of meson-meson system in
Ref.~\cite{misha} by $2m_{\Sigma_c}T(E)$ since we treat a
meson-baryon system.}

\be T(E)= \frac{1}{2m_{\Sigma_c}} \frac{-8\pi E}{p\cot
\delta(p)-i\,p}\, , \ee from where we determine the phase shifts in
the infinite volume problem.

In Fig.~\ref{Fig:pschiral}, the solid curve stands for the phase
shifts for $\pi \Sigma_c$ scattering derived from the two coupled
channels unitary approach with $\eta=1$, while the dashed line is
obtained with $\eta=0.8$. We also show with the dotted line the
phase shifts that were obtained considering only the $\pi \Sigma_c$
channel.

\subsection{The $DN$, $\pi \Sigma_c$ interaction in finite volume}

One can also use regularization with a cut off in three momentum
once the $q^0$ integration is analytically
performed~\cite{ollernpa620} with the result

\ba G_j &=& \int\limits^{|\vec q|<q_{\rm max}} \frac{d^3\vec
q}{(2\pi)^3}\frac{2m_2}{2\omega_1(\vec q)\,\omega_2(\vec q)}
\nonumber \\ && \times \frac{\omega_1(\vec q)+\omega_2(\vec q)}
{E^2-(\omega_1(\vec q)+\omega_2(\vec q))^2+i\epsilon}, \non
\omega_{1,2}(\vec q)&=&\sqrt{m_{1,2}^2+\vec q^{\,\,2}}\, ,
\label{prop_cont} \ea with $m_1$, $m_2$ corresponding to $m$ and $M$
of Eq.~(\ref{loop}) for each channel $j$. In Ref.~\cite{ollerulf}
the equivalence of the two methods was established.

When one wants to obtain the energy levels in the box, one replaces
the $G$ function by $\tilde G$, where instead of integrating over
the energy states of the infinite volume, with $q=|\,\vec{q}\,|$
being a continuous variable, as in Eq.~(\ref{prop_cont}), one sums
over the discrete momenta allowed in a finite box of side $L$ with
periodic boundary conditions. We then have $\tilde G={\rm
diag}\,(\tilde G_1,\tilde G_2)$, where \ba \tilde
G_{j}&=&\frac{1}{L^3}\sum_{\vec q}^{|\vec q|<q_{\rm max}}
\frac{2m_2}{2\omega_1(\vec q)\,\omega_2(\vec q)}\,\,
\frac{\omega_1(\vec q)+\omega_2(\vec q)} {E^2-(\omega_1(\vec
q)+\omega_2(\vec q))^2}, \non \vec q&=&\frac{2\pi}{L}\,\vec n,
\quad\vec n\in {Z}^3 , \label{tildeg} \ea with the same notation as
in Eq.~(\ref{prop_cont}).

By using the dimensional regularization of the loop function $G^D$
of Eq.~(\ref{eq:g-function}), we can write~\cite{alberto}
\begin{align}
\tilde{G}(E) &= G^D(E) \nonumber
\\ & +\lim_{q_{\rm max}\to
\infty}\Bigg[\frac{1}{L^3}\sum_{\vec q}^{|\vec q|<q_{\rm max}}I(\vec
q) -\int\limits_{q<q_{\rm max}}\frac{d^3q}{(2\pi)^3}
I(\vec q)\Bigg] \nonumber \\
&\equiv G^D(E) + B , \label{gtdim}
\end{align}
where $I(\vec q)$ is the integrand of Eq.~(\ref{prop_cont})

\ba I(\vec q)&=&\frac{2m_2}{2\omega_1(\vec q)\,\omega_2(\vec q)}
\frac{\omega_1(\vec q)+\omega_2(\vec q)} {E^2-(\omega_1(\vec
q)+\omega_2(\vec q))^2+i\epsilon} . \label{prop_contado} \ea

The three dimensional sum in Eq.~(\ref{gtdim}) can be reduced to one
dimension considering the multiplicities of the cases having the
same $\vec{n}^{\,2}$ ~\cite{mishajuelich,c}. The integral in
Eq.~(\ref{gtdim}) has an analytical form as shown in the appendix of
Ref.~\cite{d} (see erratum).

\begin{figure}[ptbh]
\begin{center}
\includegraphics[scale=0.4]{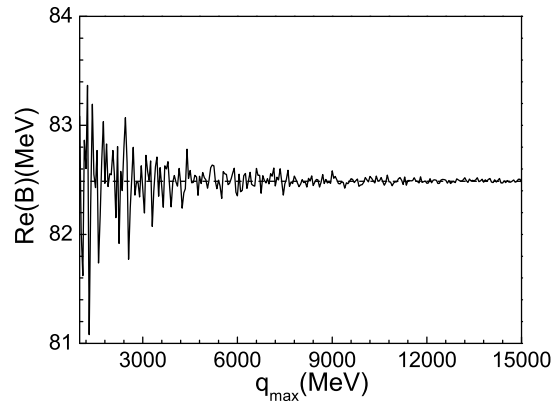} \vspace{-0.8cm}
\caption{Real part of the second part of Eq.~(\ref{gtdim}) for the
$\pi \Sigma_c$ channel. The dashed line stands for the average that
we take between $2000$ MeV and $4000$ MeV for $q_{max}$. The results
correspond to a value of $L=2.5~m_{\pi}^{-1}$ and E$= 2600$ MeV.}
\label{fig:bbar}
\end{center}
\end{figure}

In the box, the same Bethe-Salpeter equation is used substituting
$G^D$ by $\tilde G$ of Eq.~(\ref{gtdim}). When calculating the limit
of $q_{max}$ going to infinity in Eq.~(\ref{gtdim}) we obtain
oscillations which gradually vanish as $q_{max}$ goes to infinity.
Yet it is unnecessary to go to large values of $q_{max}$, and
performing an average for different $q_{max}$ values between $2000$
MeV and $4000$ MeV one obtains a perfect convergence, as one can see
in Fig. \ref{fig:bbar}.  Note that the imaginary part of $G^D$ and
that of the integral in Eq.~(\ref{gtdim}) are identical and they
cancel in the construction of $\tilde G$, which is a real function.

The eigenenergies of the box correspond to energies that produce
poles in the $T$ matrix. Thus we search for these energies by
looking for zeros of the determinant of $1-V\tilde G$ \ba
\det(1-V\tilde G) &=& 1-V_{11}\tilde G_1-V_{22} \tilde
G_2 \nonumber \\
&& +(V_{11}V_{22}-V_{12}^2)\tilde G_1\tilde G_2=0\, . \label{eq:det}
\ea

\begin{figure}[ptbh]
\begin{center}
\includegraphics[scale=0.4]{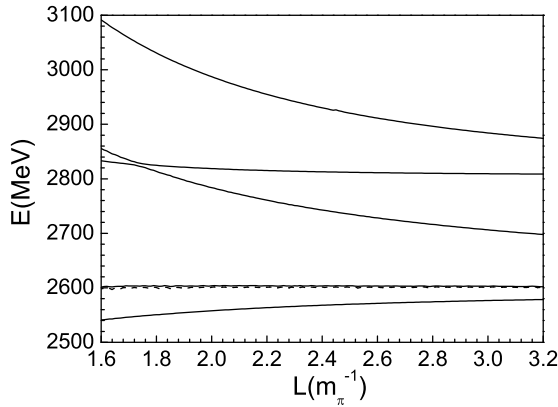}  \vspace{-0.8cm}
\caption{Energy levels as functions of the cubic box size
$L(m^{-1}_{\pi})$, derived from the chiral unitary approach.}
\label{fig:evslchiral}
\end{center}
\end{figure}

In Fig.~\ref{fig:evslchiral} we show the first five energy levels
obtained for the box for different values of $L$. We observe a
smooth behavior of the levels as a function of $L$. The second level
is special because it mainly comes from the $DN$ channel. If we only
include the $DN$ channel, we get very similar results as shown with
the dashed curve. From Fig.~\ref{fig:evslchiral} we can also see
that the second level is rather independent on the values of the
cubic box size $L$. The value for the eigenenergy of the second
level is around $2600$ MeV which is very close to the mass of
$\Lambda_c(2595)$.

\section{The inverse problem of getting phase shifts from lattice data}

In this section we face the problem of getting bound states and
phase shifts in the infinite volume from the energy levels obtained
in the box using the two channel approach of
Ref.~\cite{mizutaniprc74}, which we would consider as ``synthetic"
lattice data. To accomplish this we need more information than just
the lowest level, but we shall see that the first two levels shown
in Fig.~\ref{fig:evslchiral} already provide the necessary
information to reproduce the problem in the infinite volume.

In Ref.~\cite{misha} several methods were suggested to solve the
inverse problem, one of the methods was the following: In two
channels one has three degrees of freedom, $V_{11}$, $V_{12}$, and
$V_{22}$, or in terms of phase shifts, $\delta_1$, $\delta_2$ and
the inelasticity $\eta$ (or equivalently the mixing angle). One
strategy to obtain these magnitudes is to use three levels that
contain a certain energy and using Eq. (13) determine the three
degrees of freedom for a given energy. This strategy is used in [20]
and is also suggested in \cite{sharpe,davoudi} to obtain directly
$\delta_1$, $\delta_2$ and $\eta$. The technical problem that this
method poses is that for a given energy one might need a too large
or a too small value of $L$ that could make the computation too
lengthy or inaccurate, respectively, although using moving frames,
like in \cite{sharpe,davoudi} one obtains more levels that can
reduce the span of values of $L$. Other different methods were
suggested in Ref.~\cite{misha} and we borrow here the one based on a
fit to the data in terms of a potential parameterized as a function
of the energy suggested by the coupled channel unitary approach of
the work of Ref.~\cite{daniel} or
Refs.~\cite{Kolomeitsev:2003ac,Hofmann:2003je,Guo:2006fu}. As we can
see in Eq.~(\ref{chiralpotential}), the potentials have a large
constant part, some terms proportional to $s$ and some terms
proportional to $\sqrt{s}$. It is very easy to see that if one
chooses a region of energies around a certain value of $s$, $s_0$,
the potential can be expanded as a function of $s-s_0$ to a good
approximation. Choosing $s_0= (m_{\pi} +M_{\Sigma_c})^2$ then the
ansatz of the following equation

\be V_{ij}=a_{ij}+b_{ij}(s-(m_{\pi} +M_{\Sigma_c})^2)\, \label{fitv}
\ee is a very accurate assumption.

\begin{figure}[ptbh]
\begin{center}
\includegraphics[scale=0.4] {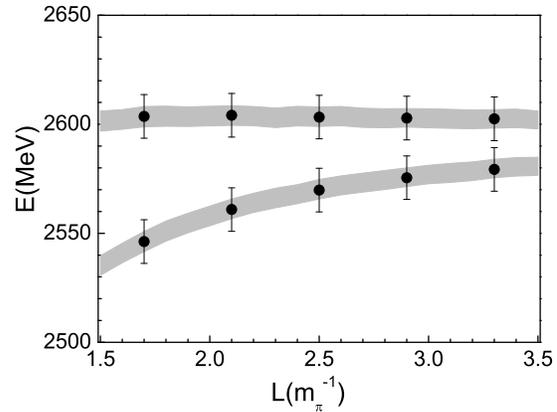}  \vspace{-0.8cm}
\caption{Energy levels as functions of the cubic box size $L$,
reconstructed from fits to the "data" of fig.~\ref{fig:evslchiral}
by using the potential of Eq.~(\ref{fitv}). The band corresponds to
the statistical errors of the fitted parameters.}
\label{Fig:fitevsl}
\end{center}
\end{figure}

We assume that the lattice studies provide us with ten eigenenergies
corresponding to the first two levels of Fig.~\ref{fig:evslchiral}
for different values of $L$ between $1.7~m_{\pi}^{-1}$ and
$3.3~m_{\pi}^{-1}$. We also assume that the levels are provided with
an error of $\pm 10$  MeV. We make a best fit to the data assuming a
potential as in Eq.~(\ref{fitv}). We look for the minimum $\chi^2$
by the MINUIT fit program and obtain a set of parameters for
$a_{ij}, b_{ij}$. The minimum $\chi^2$ that we get from the best fit
is $2.5 \times 10^{-2}$. Then we generate random sets of the
parameters within the range of error of each parameter determined by
the best fit, such that $\chi ^2$ is only increased below
$\chi_{min}^2+ 1$. With these values we generate the spectrum of
Fig.~\ref{fig:evslchiral} by searching for the zeros of the
determinant of $1-V\tilde G$. This provides a band of values for the
spectrum shown in Fig.~\ref{Fig:fitevsl}. As found in
Refs.~\cite{misha,alberto,mishajuelich,mishakappa} one has the
freedom to choose the regularization constant different to the one
used originally to generate the spectrum and the good fit to the
spectrum is obtained by a corresponding change in the parameters of
the potential, a feature tied to the renormalization group. This is
most welcome because, when the lattice data are provided to us, we
do not know which implicit regularization subtraction constant the
lattice data is supporting (the lattice spacing is not a problem in
this sense, as discussed in Ref.~\cite{mishajuelich}). The inverse
method has only a real value if the results that one obtains are
independent of this subtraction constant.

The aim of the inverse method is to get the phase shifts and bound
state in the infinite volume from the spectrum obtained in the box.
For this purpose we take now the potential obtained by the best fit
to the synthetic data with a chosen subtraction constant, and use it
in Eq.~(\ref{bse}) to produce the scattering amplitude in the
infinite volume case using $G^D$ with the same subtraction constant.

\begin{figure}[ptbh]
\begin{center}
\includegraphics[scale=0.4] {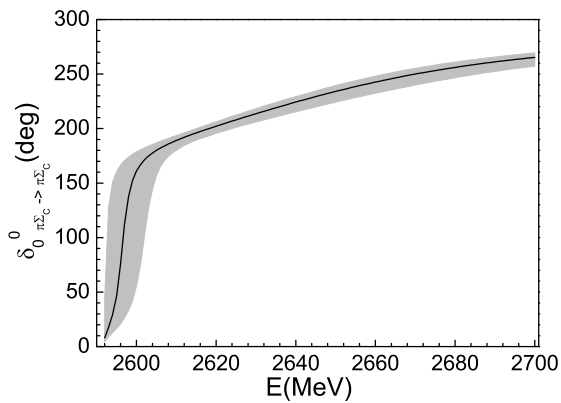}  \vspace{-0.8cm}
\caption{Phase shift for $\pi \Sigma_c$ scattering obtained from the
two coupled channels unitary approach (solid line). The band is
derived from the fits to the "data" of Fig.~\ref{Fig:fitevsl} by
using the potential of Eq.~(\ref{fitv}) with two channels.}
\label{Fig:psfitfull}%
\end{center}
\end{figure}

In Fig. \ref{Fig:psfitfull} we show the $\pi \Sigma_c \to \pi
\Sigma_c$ phase shifts, for $I=0$, $\delta^0_0$, obtained with the
uncertainty provided by the set of parameters that fulfill the
$\chi^2 < \chi^2_{\rm min}+1$ condition. As we can see, the
agreement with the exact results is quite good, and we see how the
errors in the determination of the lattice levels have propagated in
the determination of the phase shifts. The results obtained show the
presence of the $\Lambda_c(2595)$ resonance around this energy with
an uncertainty of $\pm 5$ MeV. The width that we calculate from the
position of the pole in the complex plane is about 3 MeV. One can
also obtain this from the phase shifts by using \ba T =
\frac{g^2}{E-E_R+i\Gamma/2} = \frac{1}{2m_{\Sigma_c}} \frac{-8\pi
E}{p\cot \delta(p)-i\,p} \ea which leads to the equations \ba
\frac{E-E_R}{g^2} &=&
\frac{-m_{\Sigma_c}p\cot \delta(p)}{4\pi E}, \\
\frac{\Gamma}{2g^2} &=& \frac{m_{\Sigma_c}p}{4\pi E} \ea from where
we get $\Gamma= 3.3 \pm 1.9$ MeV.

\begin{figure}[ptbh]
\begin{center}
\includegraphics[scale=0.4] {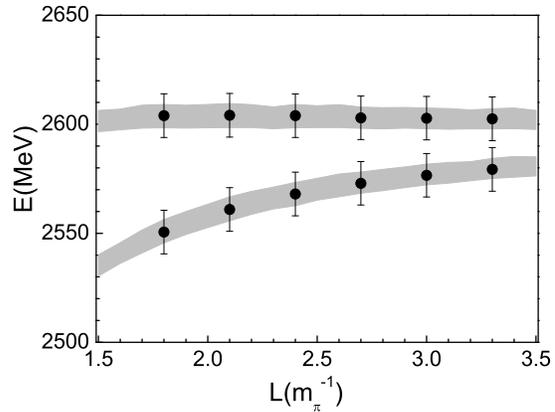}  \vspace{-0.5cm}
\caption{As in Fig.~\ref{Fig:fitevsl}, but for the new fit with the
error $\pm 5$ MeV for the centroids of the 'data'.}
\label{Fig:fitevslnew}
\end{center}
\end{figure}

At this point we want to improve the error analysis by introducing
two new ingredients. The first one is to consider that the centroids
of the data are not exactly on the exact curve, as it would
correspond to actual lattice data. For this purpose we follow
exactly the procedure done in Ref.~\cite{misha} and let the
centroids move randomly within $\pm 5$ MeV from the exact point in
the curves of Fig.~\ref{fig:evslchiral}. We then make a large number
of runs and determine the new band of results. This is shown in
Fig.~\ref{Fig:fitevslnew}. As we can see, the error band has
increased a bit, by about $20\%$. We repeat the procedure allowing
now the centroids moving randomly with $\pm10$ MeV and we find that
the band increases by an extra $20\%$ with respect to
Fig.~\ref{Fig:fitevslnew}.

\begin{figure}[ptbh]
\begin{center}
\includegraphics[scale=0.4] {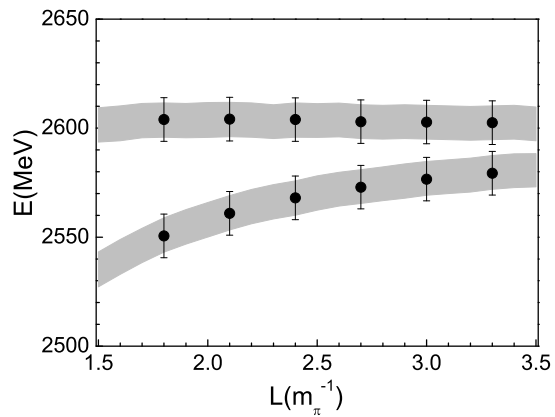}  \vspace{-0.5cm}
\caption{As in Fig.~\ref{Fig:fitevsl}, but for the new fit by using
the potential of Eq.~(\ref{fitvnew}) with two channels.}
\label{Fig:fitevslnewpotential}
\end{center}
\end{figure}

In a second step we also want to take into account the effects of
using more freedom in the parameterization of the potential. This
was discussed in Ref.~\cite{mishakappa} and was shown to be an extra
source of uncertainty. We want to investigate what happens here. For
this purpose we use a different parameterization of the potential
guided by the terms that appear in Eq.~(\ref{chiralpotential}).
Thus, we take

\begin{eqnarray}
V_{ij} = && a_{ij} + b_{ij} (\sqrt{s}-(m_{\pi}+M_{\Sigma_c})) \nonumber \\
&& + c_{ij} (s-(m_{\pi}+M_{\Sigma_c})^2), \label{fitvnew}
\end{eqnarray}
hence, introducing three more parameters. We should note that with
such large number of parameters one will not only be producing a
smooth fit to the data, but the fit will also search situations to
reproduce the fluctuations. This means we should consider the extra
errors that come from this source certainly as an upper bound. The
new results, considering also the dispersion of the centroids, are
shown in Fig.~\ref{Fig:fitevslnewpotential} (note that we added also
a new data point). As we can see, and similarly to what was fond in
Ref.~\cite{mishakappa}, the error band has increased and now extends
over the whole range of the assumed errors of the lattice data.

\begin{figure}[ptbh]
\begin{center}
\includegraphics[scale=0.4] {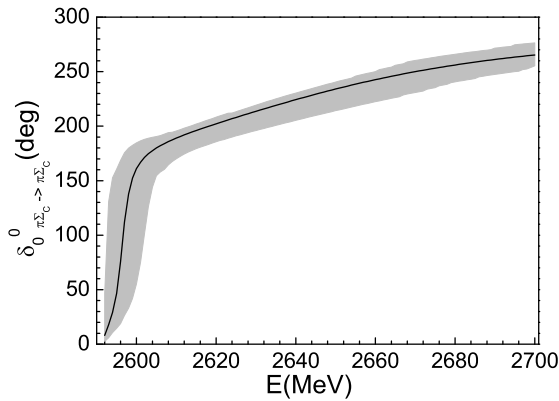}  \vspace{-0.8cm}
\caption{As in Fig.~\ref{Fig:psfitfull}, but the band is derived
from the fits to the "data" of Fig.~\ref{Fig:fitevslnewpotential} by
using the potential of Eq.~(\ref{fitvnew}) with two channels.}
\label{Fig:psfitfull3para}%
\end{center}
\end{figure}

With the new potential we reevaluate the phase shifts with two
channels as we have done for the Fig.~\ref{Fig:psfitfull}, and the
new results are shown in Fig.~\ref{Fig:psfitfull3para}. The
agreement with the exact results is good but the error band is now
increased.

The analysis done here is based on the dominance of the $\pi \Sigma_
c$ and $DN$ two body channels. The inclusion of three body channels
makes the work technically much more involved, but has already been
addressed formally in Ref.~\cite{akakaitres}. In the present case
the $\Lambda_c(2595)$ resonance can decay into $\pi \Sigma_ c$ but
also in uncorrelated $\Lambda_c \pi \pi$. Fortunately the branching
ratio in this channel is small and with large uncertainties, $18 \pm
10\%$~\cite{pdg}, that justifies its neglect in the present work.
Should there be better data in the future on this branching ratio,
and should one aim at a very accurate solution, some work along the
lines of Ref.~\cite{akakaitres} would be advisable.

\section{One channel analysis}

Since the $\pi \Sigma_c$ channel is the only one open, one might be
tempted to apply L\"uscher's approach with just one channel. One
would be assuming implicitly that the effect of the $DN$ channel
would be absorbed in the $\pi \Sigma_c$ potential. In such a case,
the energy spectrum of Fig.~\ref{Fig:fitevsl} would be given by the
poles of \ba T(E) = \frac{1}{V^{-1}-\tilde{G}(E)} \ea which gives us
$V^{-1}=\tilde{G}$ for a value of $E$ eigenenergy of the box. For
this energy, we can then write the scattering amplitude in infinite
volume as \ba T(E) = \frac{1}{\tilde{G}(E) - G(E)}. \ea

However, the direct application of this formula does not give
information above the $\pi \Sigma_c$ threshold from the first level,
since the eigenenergy values of the first level are below the $\pi
\Sigma_c$ threshold. Besides, the second level is very stable and
only gives us an energy point, with errors in the energy bigger than
the width of the resonance. It is not possible to reconstruct the
phase shifts in these circumstances. Because of this, it is more
appropriate to consider all the data of one level, since one is then
using the information on the correlation of these data. Then we fit
all these data (five points) with the potential of Eq.~(\ref{fitv})
but with only the $\pi \Sigma_c$ channel (2 parameters). We get a
good fit to the first level with $\chi^2_{min}=3.4\times 10^{-2}$.
Then we use this potential to determine the infinite volume
scattering amplitude $T$ \be T(E) = \frac{1}{V^{-1}-G(E)}.
\label{teinfinite} \ee

\begin{figure}[ptbh]
\begin{center}
\includegraphics[scale=0.4] {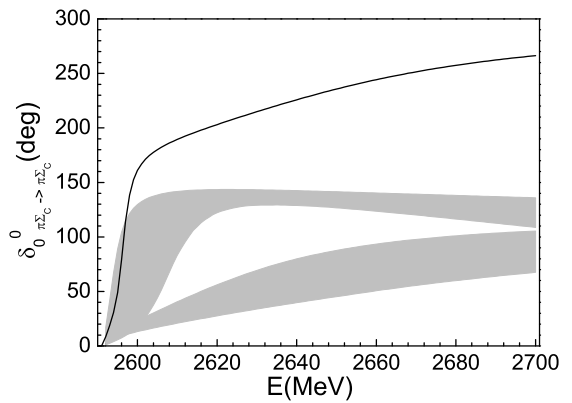}  \vspace{-0.8cm}
\caption{Phase shift $\delta^0_0$ for $\pi \Sigma_c$ scattering.
Solid line: exact results obtained from the two coupled channels
unitary approach. Lower band: derived from the fits to the first
level "data" of Fig.~\ref{Fig:fitevsl} by using the potential of
Eq.~(\ref{fitv}) with only one $\pi \Sigma_c$ channel. Upper band:
derived from the fits to the second level "data" of
Fig.~\ref{Fig:fitevsl} by using the potential of Eq.~(\ref{fitv})
with only one $\pi \Sigma_c$ channel.}
\label{Fig:fitps1channel}%
\end{center}
\end{figure}

The results for $\pi \Sigma_c$ phase shifts, with the lower band,
are shown in Fig.~\ref{Fig:fitps1channel}. As we can see, the
results obtained differ substantially from the exact results
obtained with the two coupled channels, as shown in
Fig.~\ref{Fig:fitps1channel} by the solid line. We should note that,
not even the scattering length is provided correctly.

Similarly we take now the second level in Fig.~\ref{Fig:fitevsl} and
perform the same exercise. The best fit gives a larger
$\chi^2_{min}=0.4$. The phase shifts obtained are also shown in
Fig.~\ref{Fig:fitps1channel} with the upper band. They are also in
very bad agreement with the real results.

Besides, we can take the second level of Fig.~\ref{Fig:fitevsl} and
analyze it with just the $DN$ channel. Then we cannot get the $\pi
\Sigma_c$ phase shifts, but we see that the scattering amplitude
obtained in the infinite volume by using Eq.~(\ref{teinfinite}) has
a pole at $E=2601 \pm 6$ MeV, corresponding to a $DN$ bound state.
This is telling us that the second level of Fig.~\ref{Fig:fitevsl}
is mostly tied to the $DN$ channel. One could guess that the
stability of the energy level as a function of $L$ is showing the
presence of a bound state, although this is not always the case, as
shown in Ref.~\cite{alberto}. Yet, in the present case we have gone
one step beyond, because our inverse analysis provides a parameter
set for the potential from where we can determine poles, couplings
and then test Eq.~(\ref{eq76}) for the compositeness condition. In
the present case this renders a value of around $0.9$ for
$-g^2_{DN}\frac{dG_{DN}}{dE}$, from where we conclude that the state
corresponds essentially to a $DN$ bound state, weakly decaying into
$\pi \Sigma_c$.

\begin{figure}[ptbh]
\begin{center}
\includegraphics[scale=0.4] {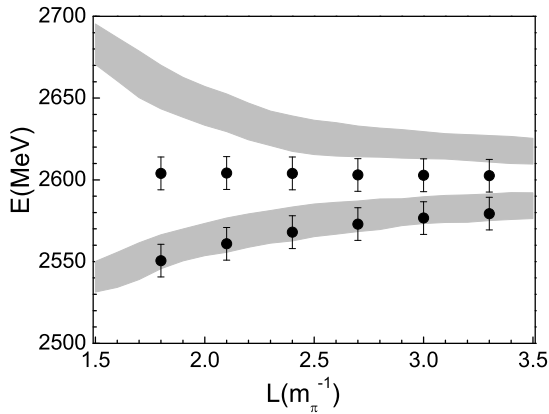}  \vspace{-0.5cm}
\caption{As in Fig.~\ref{Fig:fitevsl}, but for the new fit with the
error $\pm 5$ MeV for the centroids of the 'data' and by using the
potential of Eq.~(\ref{fitvnew}) with only one $\pi \Sigma_c$
channel.} \label{Fig:fitevsl1channel3para}
\end{center}
\end{figure}

Once again we repeat the exercise done at the end of the former
section and take into account at the same time the two effects
considered here. Thus, we consider the dispersion of the centriods
of the data and use the new parametrization of Eq.~(\ref{fitvnew}).
In this case we have now three parameters in the potential. The new
results can be seen in Fig.~\ref{Fig:fitevsl1channel3para}. We can
see that the quality of the fit to the two levels is worse than the
one obtained in Fig.~\ref{Fig:fitevsl} with two channels.

\begin{figure}[ptbh]
\begin{center}
\includegraphics[scale=0.4] {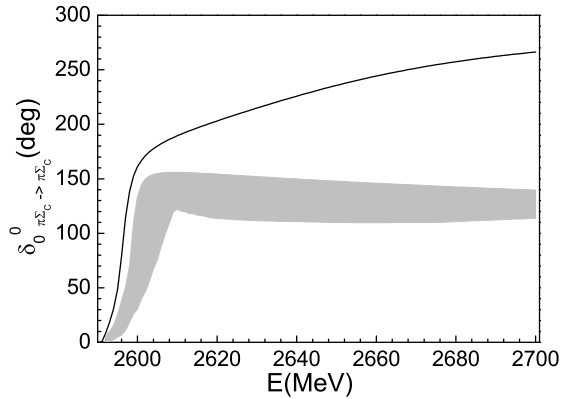}  \vspace{-0.5cm}
\caption{As in Fig.~\ref{Fig:fitps1channel}, but, the band is
derived from the fits to the first two levels "data" of
Fig.~\ref{Fig:fitevsl1channel3para} by using the potential of
Eq.~(\ref{fitvnew}) with only one $\pi \Sigma_c$ channel.}
\label{Fig:psfit3para}%
\end{center}
\end{figure}

Next, we evaluate the phase shifts with this new potential, fitted
to the two levels, and the results can be seen in
Fig.~\ref{Fig:psfit3para}. Once again, we see that assuming just the
$\pi \Sigma_c$ channel in the analysis leads to unrealistic $\pi
\Sigma_c$ phase shifts.

In summary, we see the problems that arise when we try to use
L\"uscher's approach for the interpretation of the lattice spectrum
in a case where the relevance of a closed channel is huge, like in
the present case. We also see that doing the fit analysis, that
allows us to circumvent L\"uscher's approach, but using only one
channel also fails to provide realistic phase shifts. The analysis
with two channels shows here the tremendous power that the coupled
channel approach has in this case.

\section{A test in terms of a CDD potential}

On the other hand, there is the possibility that the nature of
$\Lambda_c(2595)$ resonance could be a genuine state, not
dynamically generated by the $DN$ interaction. For this purpose we
have made a test introducing a different potential for the $DN$
interaction where a CDD pole (Castillejo, Dalitz, Dyson)~\cite{cdd}
is introduced by hand. The potential for $DN$ interaction now is,

\begin{eqnarray}
V=V_M+\frac{g^2_{CDD}}{\sqrt{s}-\sqrt{s_{CDD}}},
\label{potentialcdd}
\end{eqnarray}
where $V_M$ is assumed to be energy independent and $g^2_{CDD}$,
$s_{CDD}$ are the parameters of the CDD pole.

Like it has been done in Ref.~\cite{alberto}, from the above
potential, we now find

\begin{eqnarray}
1-\frac{g^2_{CDD}G^2}{(\sqrt{s}-\sqrt{s_{CDD}})^2\frac{dG}{dE}} =
\frac{1}{1-Z}, \label{1minusZ}
\end{eqnarray}
with $Z$ the field renormalization constant for the genuine state,
which accounts for the probability to have a genuine state.

We take the potential Eq.~(\ref{potentialcdd}) with $V_M$ of the
order of $10$ times smaller that the potential used for $V_{22}$,
$\sqrt{s_{CDD}}$ corresponding to a 20 MeV below the mass of the
$\Lambda_{c}(2595)$ and then $g^2_{CDD}=1.86$ such  as to get the
bound state at the mass of $\Lambda_c(2595)$. We find that at the
pole of this state, from Eq.~(\ref{1minusZ}), we get $Z=0.96$, which
shows that the introduction of a CDD pole as in in
Eq.~(\ref{potentialcdd}) is good enough to generate a genuine state.

\begin{figure}[ptbh]
\begin{center}
\includegraphics[scale=0.4]{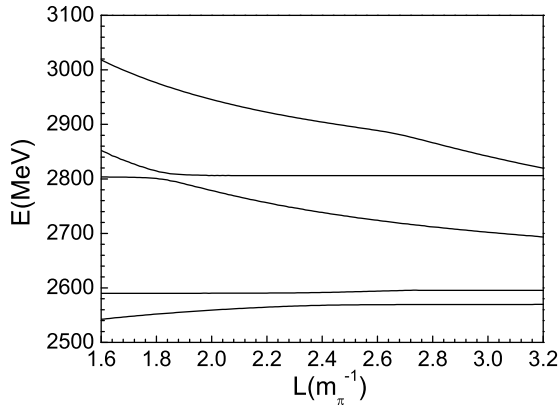}  \vspace{-0.8cm}
\caption{Energy levels as functions of the cubic box size
$L(m^{-1}_{\pi})$, derived from the potential of
Eq.~(\ref{potentialcdd}) for the $DN$ channel.} \label{fig:evslcdd}
\end{center}
\end{figure}

The energy levels in the box with the potential of
Eq.~(\ref{potentialcdd}) for the $DN$ channel are shown in
Fig.~\ref{fig:evslcdd}. As we can see, the levels are different from
those obtained with the couple channel potential, which are shown in
Fig.~\ref{fig:evslchiral}. It is clear that the determination of the
levels with lattice calculations can differentiate between the two
different types for the potentials.

Next, we perform a fit, with a CDD potential
Eq.~(\ref{potentialcdd}) for $V_{22}$, to the first two levels of
Fig.~\ref{fig:evslchiral} that are obtained with the couple channel
potential. We have taken also ten points over the curves and assumed
$\pm 10$ MeV errors as we have done above. The best fitting results
are: $g_{CDD}= 91.4 \pm 26.9$, $\sqrt{s_{CDD}}= 8190 \pm 1600$ MeV,
and $V_M = 1.23 \pm 0.45$, with these values and their
uncertainties, we can get now, at $E=2596$ MeV, $Z=0.15 \pm 0.07$
from Eq.~(\ref{1minusZ}), which is in the order of $15\%$ with large
error. This means around $85\%$ fraction of $\Lambda_c(2595)$ being
dynamically generated. We should note that although we get a good
fit with a potential that formally contains a CDD pole, the large
value of the mass of the CDD pole renders the potential smooth, as
in the case of the coupled channels, and the test tells us that the
state corresponds to a dynamically generated one. We should also
mention that we now only take two energy levels of
Fig.~\ref{fig:evslchiral} for fitting, this is why the best results
have large errors. If we took more energy levels, we could determine
these values with more precision. But our present result, $Z = 0.15
\pm 0.07$, suffices to show that from this limited information one
can get valuable conclusions on the nature of the $\Lambda_c(2595)$
resonance, which is mostly a $DN$ bound state.

\section{Summary}

In this work, we study the interaction of the coupled channels $DN$
and $\pi \Sigma_c$ in an SU(4) extrapolation of the chiral unitary
theory. The resulting interaction is used to reproduce the position
of the $\Lambda_c(2595)$ resonance in the isospin zero $DN$ channel.
Then we conclude that the $\Lambda_c(2595)$ is mostly a $DN$ bound
state.

We then study the interaction of the coupled channels $DN$ and $\pi
\Sigma_c$ in the finite volume. Energy levels in the finite box are
evaluated. We assume that the results obtained would correspond to
results given by lattice calculations. From there we address the
inverse problem. We propose a rather general and realistic potential
and, using two coupled channels, a fit to the synthetic data is made
assuming some reasonable errors in the data. Then this potential  is
used in the infinite volume case, generating the $\pi \Sigma_c$
phase shifts within an error band around the original results. This
part provides information  for lattice QCD calculations about the
accuracy in the energies of the spectrum needed to get a desired
accuracy in the phase shifts.

A second part of the investigation was about the use of a one
channel L\"uscher's approach, with just the open $\pi \Sigma_c$
channel, to induce $\pi \Sigma_c$ phase shifts from the finite
volume spectrum. We found in this case that, due to the large weight
of the closed $DN$ channel in this problem, the results obtained
using L\"uscher's approach with just the $\pi \Sigma_c$ channel was
of no use. Even more, making a fit analysis to the lattice data with
just the $\pi \Sigma_c$ channel produced erroneous $\pi \Sigma_c$
phase shifts. Certainly one does not know a priori from the lattice
QCD results whether two channels would be necessary in the analysis.
However, we also showed that the results obtained from the analysis
of the first two levels with just one channel were different to each
other. This could be taken as a clear indication that at least two
channels are needed in a realistic analysis of the lattice QCD
results in such a case. The results from the chiral unitary
approach, and the two channel formalism shown here, which can be
trivially generalized to more channels, provide a good perspective
to undertake future lattice QCD calculation in this sector. We also
showed that the analysis done here, not only provides us with the
$\pi \Sigma_c$ phase shifts and the presence of a bound state, but
through the test of the sum rule of Eq.~(\ref{eq76}) (essentially
Weinberg's compositeness test), it also tells us that this bound
state corresponds to a molecular state of a $DN$ system.

\section*{Acknowledgments}

We would like to thank J. Nieves and M. D\"oring  for useful
discussions. This work is partly supported by DGICYT Contract No.
FIS2006-03438, the Generalitat Valenciana in the project PROMETEO,
the Spanish Consolider Ingenio 2010 Program CPAN (CSD2007-00042) and
the EU Integrated Infrastructure Initiative Hadron Physics Project
under contract RII3-CT-2004-506078 and by the National Natural
Science Foundation of China (NSFC) under grant n. 11105126.

\end{document}